 \documentclass[aps]{revtex4}

 \usepackage{graphicx}

\usepackage{amssymb}

\begin{document}

\title{Forcing reversibility in the no strand-bias substitution model
allows for the theoretical and practical identifiability of its
5 parameters from pairwise DNA sequence comparisons.}

\author{Osvaldo Zagordi}
\email{zagordi@sissa.it}
\affiliation{International School of Advanced Studies SISSA-ISAS\\ via Beirut 2-4, 34013 Trieste, Italy}

\author{Jean R. Lobry}

\affiliation{Laboratoire BBE-CNRS-UMR-5558, Univ. C. Bernard - Lyon I\\
43 Bd 11/11/1918, F-69622 Villeurbanne CEDEX, France}

\begin{abstract}
Because of the base pairing rules in DNA, some mutations experienced by a portion of DNA during its
evolution result in the same substitution, as we can only observe differences in coupled nucleotides.
Then, in the absence of a bias between the two DNA strands, a model with at most
6 different parameters instead of 12 is sufficient to study the evolutionary relationship between homologous sequences derived from a common
ancestor. On the other hand the same symmetry reduces the number of independent observations which can be made. Such a reduction
can in some cases invalidate the calculation of the parameters. A compromise between biologically acceptable hypotheses and tractability
is introduced and a five parameter \textit{reversible no-strand-bias condition} (\textbf{RNSB}) is presented.
The identifiability of the parameters under this model is shown by examples.
\end{abstract}



\maketitle


\section{\label{intro}Introduction}
Darwinian Evolution is based upon the interplay of two driving forces: \textbf{mutation} of an organism features, and \textbf{natural 
selection} acting on the living organisms. Nowadays the role of the DNA in the evolutive processes has been recognised, and the physical 
basis of the mutation process has been identified. Mutation acts on the DNA and we call \textsl{mutation rate} the probability that a
descendant has a difference in the genome if this is compared to that of its parents.
The substitution rate is the probability of finding a difference when comparing the genomes of species to one of its ancestors.

We see that while the mutation is closely related to the biophysical process of DNA damage, or replication error etc., the substitution
is the result of a mutation and of a population-dynamics process, which has spread the former to the whole population.
A fundamental observation by M. Kimura in 1968 \cite{ki68} argued that, in the case of neutral mutations (i.e. those mutations which have
no apparent effect on the adaptation of an organism to the environment), we can deduce the mutation rate from the substitutions, as
they are actually the same.

Let's consider an ancestor $O$ at time $t=0$ which separates into two different evolutive lineages, 
resulting in two different species, $A$ and $B$ at time $t$. 
It would be useful to define a distance between $A$ and $B$ and to have a tool to calculate it by just comparing the 
genomes of $A$ and $B$.

In order to study evolutionary distance between homologous DNA sequences (descending from a common ancestor) and their
consequent relationship, a model for nucleotide substitution can be introduced.
Generally, the process is assumed to be a Markov chain, if some assumptions are made about the underlying process.
The general hypotheses are:
\begin{itemize}
\item substitution rates do not depend on the position along the DNA sequence;
\item they are constant during evolutionary time;
\item the two evolutionary lineages have the same rates;
\item DNA sequences are at the compositional equilibrium when they start to diverge (nucleotide frequencies are constant).
\end{itemize}
We will see that even with relaxing the last two hypotheses some calculations can be performed, but
it is worth noting that compositional equilibrium, if the last assumption is verified, is maintained during the course of evolution.

Denoting with $f_{i}$ the compositional equilibrium frequency of the nucleotide $i$ with $i \in  \{ \sf{A, T, G, C} \}$
and with $r_{ij}=r_{i\leftarrow j}$ the substitution rate from nucleotide $j$ to $i$ in the unit time.
The distance between two sequences, can now be defined as 

\begin{equation}
d=2t\sum_{i}f_{i}\mu_{i}=2t\sum_{i}f_{i}\sum_{j(\neq i)}r_{ji} \quad .
\label{distance}
\end{equation}

\begin{figure}[c]
\begin{center}
\includegraphics[width=7cm]{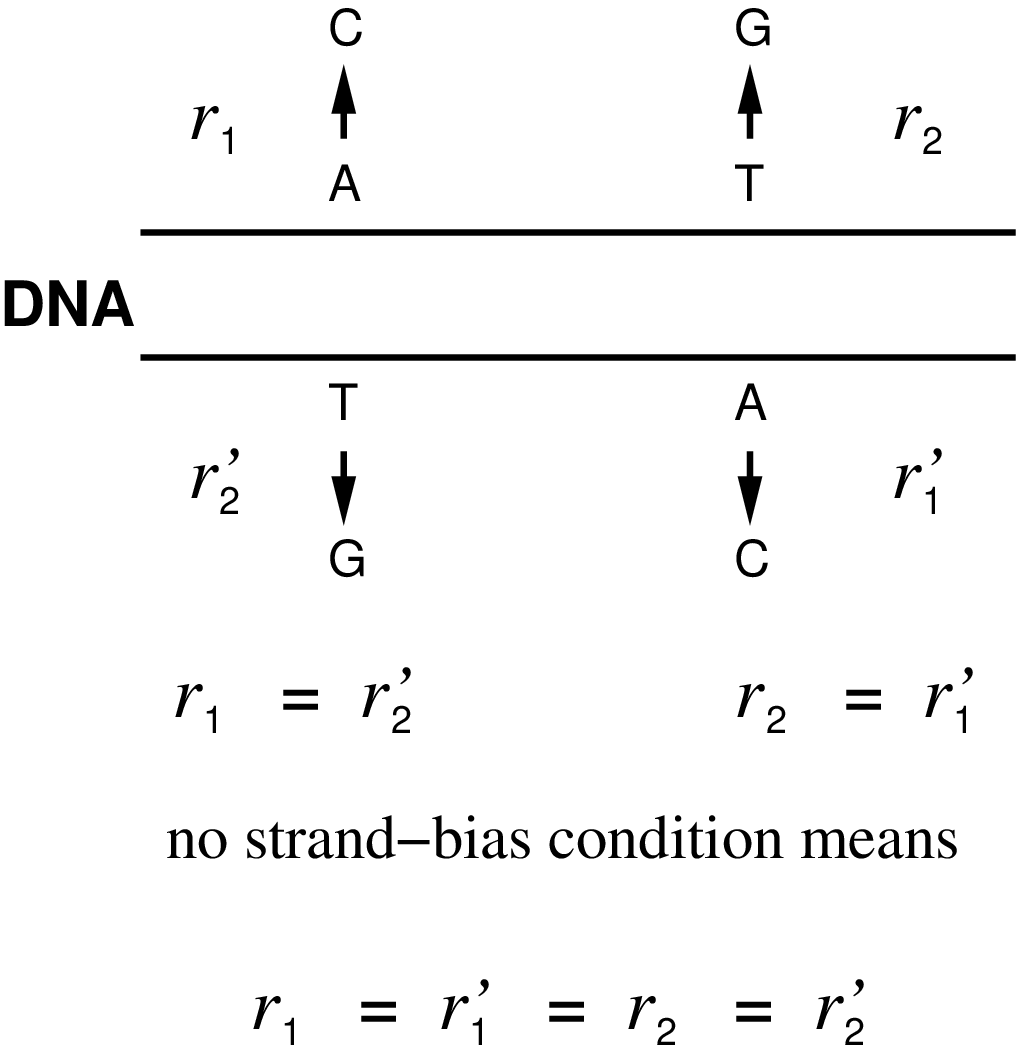}
\caption{\footnotesize{Explication of the \textsl{no strand-bias condition}. If the rates for a certain substitution are the same on 
both strands of DNA, one can deduce the equivalence of this rate to the one between the complementary bases.}}\label{nsbc_image}
\end{center}
\end{figure}

\begin{figure}[c]
\begin{center}
\includegraphics[width=7cm]{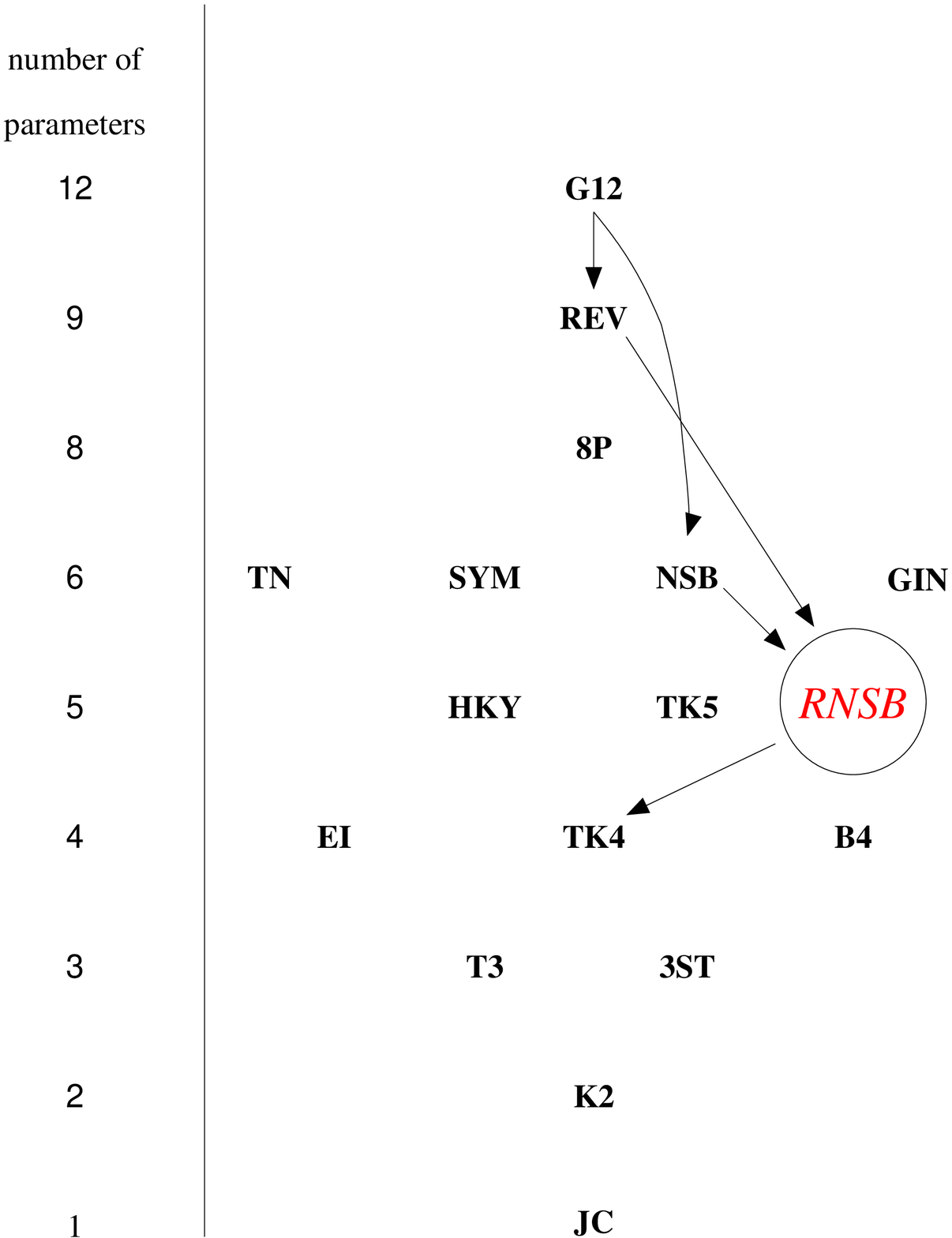}
\caption{\footnotesize{Hierarchy of DNA substitution models. Simplifications leading from a model to a simpler one are indicated by arrows.
Only those directly referring to our discussion are drawn. This figure has been adapted from Robert Schmidt's work.}}\label{schema}
\end{center}
\end{figure}

Since 1969, when Jukes and Cantor proposed their first one-parameter model for nucleotide subsitution in DNA, many different models of
increasing complexity have been published. The general 4-state Markov model has 12 independent parameters, \textbf{G12} in
fig.\ref{schema} (for a review see Zharkikh \cite{zh94}). This number, and consequently the
model complexity, can be decreased by further conditions on the parameters, leading to a plethora of different models. A possible choice is
to take into account the property of \textsl{no strand-bias}, explained in fig.\ref{nsbc_image}.
It was introduced by Sueoka in 1995 \cite{su95} and we generally refer to it as
\textit{type 1 parity rule} or \textit{PR1}. This rule is easily understood thinking that, scoring the substitution on one strand,
the same substitution can be obtained in two ways: $\sf{A} \rightarrow \sf{C}$ is observed also if on the opposite strand
$\sf{T} \rightarrow \sf{G}$.

This means that we cannot discriminate substitutions between two bases from those between their complementary bases. In symbols:
\begin{equation}
r_{ij}=r_{\bar{\imath}\bar{\jmath}},
\end{equation}
where the bar means complementary nucleotide: $\bar\mathsf{A}=\mathsf{T}$ and viceversa. And $\bar\mathsf{C}=\mathsf{G}$ similarly.

The number of independent parameters is then halved, so that the following substitution rates can be introduced:

\begin{eqnarray}\label{rates}
a&\equiv& r_{\sf{AT}}=r_{\sf{TA}}\nonumber\\
b&\equiv& r_{\sf{AG}}=r_{\sf{TC}}\nonumber\\
c&\equiv& r_{\sf{CT}}=r_{\sf{GA}}\nonumber\\
d&\equiv& r_{\sf{AC}}=r_{\sf{TG}}\nonumber\\
e&\equiv& r_{\sf{CA}}=r_{\sf{GT}}\nonumber\\
f&\equiv& r_{\sf{CG}}=r_{\sf{GC}}.\nonumber
\end{eqnarray}

The notation introduced here is consistent with the one previously used by Sueoka \cite{su95} and Lobry \cite{lo95}

Equilibrium frequencies for such a model are easily derived from the \textsl{master equations}:
$$
\dot{q_{i}}=\sum_{j}(r_{ij}q_{j}-r_{ji}q_{i}), 
$$
where $q_{i}$ denotes in general the probability of state $i$.

These frequencies are given by:
\begin{eqnarray}\label{equil}
f_{1} & \equiv & q^{\infty}_{\sf{A}}=q^{\infty}_{\sf{T}}=\frac{1}{2}\frac{b+d}{b+c+d+e}\nonumber\\
\mbox{}\\
f_{2} & \equiv & q^{\infty}_{\sf{G}}=q^{\infty}_{\sf{C}}=\frac{1}{2}\frac{c+e}{b+c+d+e}.\nonumber
\end{eqnarray}
The intrinsic symmetry of the model is evident. In this framework, in other words, there is only \textbf{one} independent frequency, the
other being deduced by the normalization condition $2f_{1}+2f_{2}=1$.
We now stress the fact that this is valid in a single strand (\textit{type 2 parity rule} or \textit{PR2}). If \textit{PR1} is
satisfied, then as a consequence the frequency of a nucleotide in a strand must be equal to that of its complement in the same strand.

In the following we will resume some general results regarding \textit{PR1} algebra showing that, in many cases, it is not possible to
reconstruct the supposed underlying mutation pattern because the independent parameters outnumber the possible independent observations.

\section{\label{pr1}Materials and Methods}
In this section we will give some results regarding the model introduced above, focusing on the number of actual independent
possible observations.

\subsection{\label{general}General model}

Given the substitution matrix $\mathsf{R}_{[4,4]}$, whose entries are the mutation rates per nucleotide per unit of time, 
one can deduce the \textsl{evolutionary matrix} $\mathsf{P}_{[4,4]}(t)$,
whose entries $p_{ij}(t)$
represent the probability of finding at a certain site the base $i$ at time $t$, given the base $j$ at $t=0$. Yet the 
\textsl{divergence matrix} $\mathsf{X}_{[4,4]}(t)$ can be deduced, whose entries $x_{ij}(t)$ are the mutual probability of 
finding at time $t$ the base $j$ in a sequence, given the base $i$ at the same site of the other sequence. 
Obviously, if the substitution pattern is the same for both sequences, it results in $x_{ij}(t)=x_{ji}(t)$. 

It is worth noting that the divergence matrix at initial time is nothing but the diagonal matrix with nucleotide 
frequencies on the diagonal.

The result of an evolutive process can be synthetically represented as an initial diagonal divergence matrix,
multiplied on the left and on the right by a certain number of substitution matrices (corresponding to the generation steps
in the two evolution lineages), producing a final matrix

\begin{eqnarray}
\mathsf{X}(t) & = & \mathsf{R'_m}\cdots\mathsf{R'_2}\mathsf{R'_1}~\mathsf{X}(0)~\mathsf{R^{t}_1}\mathsf{R^{t}_2}\cdots\mathsf{R^{t}_n}\nonumber\\
\mathsf{X}(t) & = & \mathsf{P'}~\mathsf{X}(0)~\mathsf{P^{t}}\label{discretex}\\
x_{ij}(t) & = & \sum_{k=1}^{4}p'_{ik}(t)f_{k}p_{jk}(t)\nonumber
\end{eqnarray}
where the substitution matrices can, in principle, all be different.

The entries of the divergence matrix are the experimentally observable quantities. 

In our case the substitution matrix is $\mathsf{R}_{[4,4]}$:

\begin{displaymath}
\begin{array}{|c|c|c|c|c|}
\hline
\Rsh    & \sf{A}                    & \sf{T}                   & \sf{G}                   & \sf{C}\\
\hline
\sf{A}  & 1-a-c-e & a                   & c                   & e                   \\
\hline
\sf{T}  & a                   & 1-a-e-c & e                   & c                     \\
\hline
\sf{G}  & b                   & d                   & 1-b-d-f & f                       \\
\hline
\sf{C}  & d                   & b                   & f                   & 1- d - b - f  \\
\hline
\end{array}
\end{displaymath}

obtained under the hypotheses of \textit{no-strand-bias}, I.E. \textit{PR1}.

\subsection{Non identifiability of some models}
In the following we show that the mathematical properties of the \textit{PR1} algebra are such that,
dealing with the general model, the parameters to estimate outnumber the possible independent observations, so that the model
is untractable.
As seen in eq.(\ref{discretex})
$$
\mathsf{X}(t) = \mathsf{P'}~\mathsf{X}(0)~\mathsf{P^{t}}.
$$
Now, several cases are possible, depending on whether $\mathsf{P'}=\mathsf{P}$ or not.
In the following, we will assume that $\mathsf{X}(0)$ is already at compositional equilibrium, I.E.
\begin{eqnarray}
q^{0}_{\sf{A}} = & q^{0}_{\sf{T}} = f_1 = & x_{AA}(t=0) = x_{TT}(t=0) \nonumber\\
\mbox{}\\
q^{0}_{\sf{C}} = & q^{0}_{\sf{G}} = f_2 = & x_{CC}(t=0) = x_{GG}(t=0) \nonumber
\end{eqnarray}

\subsubsection{$\mathsf{P'}=\mathsf{P}$} \label{counting}
As $\mathsf{P'}=\mathsf{P}$ it is clear that  $\mathsf{X}(t)$ is symmetric ($\mathsf{X}(t) = \mathsf{X^{t}}(t)$).
We have to estimate 6 parameters (6 mutation rates) and we have only 5 independent observations.
This happens because of the symmetry $x_{ij}=x_{ji}$,
the normalization conditions and because $x_{ij}=x_{\bar{\imath}\bar{\jmath}}$.
In more detail:
\begin{eqnarray}\label{4par1}
x_{AG} & = & x_{GA}=x_{TC}=x_{CT}\nonumber\\
x_{AC} & = & x_{CA}=x_{TG}=x_{GT}\nonumber\\
x_{AT} & = & x_{TA}\nonumber\\
x_{CG} & = & x_{GC}\nonumber\\
x_{AA} & = & x_{TT}\nonumber\\
x_{CC} & = & x_{GG}\nonumber
\end{eqnarray}
Where $x_{AA} = x_{TT}$ and $x_{CC} = x_{GG}$ can be deduced by the other four using the normalization ($\sum_{j}x_{ij}=f_{i}$) 
and the equilibrium frequencies.
We find that $x_{AG}$, $x_{AC}$, $x_{AT}$, $x_{CG}$ and one equilibrium frequency are the only independent observable quantities.

\subsubsection{$\mathsf{P'} \neq \mathsf{P}$}

In this case mutation rates double becoming 12;
so we have 12 parameters to calculate. Independent observations, on the other hand, increase up to 7, because of the lack of the
symmetry $x_{ij}=x_{ji}$. Still the model is intractable.

\subsection{\label{case}Reversible \textit{PR1} model}
In this section we will deal with one of the previous models, the simplest one where $\mathsf{P'}=\mathsf{P}$.
In this case simple calculations lead to an analytical expression for the divergence matrix, but the model
remains intractable. Yet we will see that by the imposition of a certain property the model becomes tractable, and a way to
estimate the parameters for a real data set will be proposed.

In the following we will assume again that the initial divergence matrix is already at compositional equilibrium.
Further, we will treat the evolutionary process as a continuous time process, being the time since the divergence very long.
This allows us to write the following equations to solve the problem.
The expression for the evolutionary matrix is
\begin{equation}
\mathsf{P}(t)=\exp\{\mathsf{R}t\};
\label{pdt}
\end{equation}
as it is the solution of the differential equations (see Rodriguez et al. \cite{ro90})

\begin{eqnarray}
\frac{d\mathsf{P}(t)}{dt} & = & \mathsf{P}(t)\mathsf{R}\\
\frac{dp_{ij}(t)}{dt} & = & \sum_{k=1}^{4}p_{ik}(t)r_{kj}.
\label{dpdt}
\end{eqnarray}

While the divergence matrix is given by

\begin{eqnarray}
\mathsf{X}(t) & = & \mathsf{P'}(t)\mathsf{X}(t=0)\mathsf{P}^{T}(t)\\
x_{ij}(t) & = & \sum_{k=1}^{4}p'_{ik}(t)f_{k}p_{jk}(t);
\label{xdt}
\end{eqnarray}

It is easily verified that, if $\mathsf{P'}=\mathsf{P}$, then $x_{ij}(t)=x_{ji}(t)$.

Now, the expressions for $x_{ij}(t)$ (the observables) can be inverted to obtain the rates and then the distance.

The strategy could be:
\begin{itemize}
\item solve the model, that is find the $x_{ij}(t)$ as a function of rates;
\item invert the above equations to get an expression for the rates;
\item substitute the observed quantities $\bar{x}_{ij}$ in order to have a numerical estimation of the rates;
\item use these estimates to obtain the distance.
\end{itemize}

The expressions for $x_{ij}$ can be deduced in a manner analogous to that proposed by Takahata \&  Kimura in 1981 \cite{tk81}
who deal with a slightly less general model than this (model \textbf{TK5} in fig.\ref{schema}). 
In this way we get an expression for every entry of the divergence matrix, but with five
independent expressions, as stated above. We repeat here the reasons:
\begin{itemize}
\item the symmetry of the matrix $x_{ij}=x_{ji}$;
\item the intrinsic symmetry of the model $x_{ij}=x_{\bar{\imath}\bar{\jmath}}$;
\item the normalization conditions $\sum_{j}x_{ij}=f_{i}$.
\end{itemize}
Thus, we can write down the entire divergence matrix by means of the following quantities:

\begin{eqnarray}\label{4par}
P~ & \equiv & x_{AG}=x_{GA}=x_{TC}=x_{CT}\nonumber\\
R~ & \equiv & x_{AC}=x_{CA}=x_{TG}=x_{GT}\nonumber\\
Q_{1} & \equiv & x_{AT}=x_{TA}\nonumber\\
Q_{2} & \equiv & x_{CG}=x_{GC},\nonumber\\
S_{1} & \equiv & x_{AA}=x_{TT}\nonumber\\
S_{2} & \equiv & x_{CC}=x_{GG}\nonumber
\end{eqnarray}
Where, as stated above, $S_{1}$ and $S_{2}$ can be deduced by the other four using the normalization and the equilibrium frequencies.
We find that $P, R, Q_{1},Q_{2}$ and one equilibrium frequency are the only independent observable quantities.

\subsection{Solution of the model}
Deriving an analytical expression for the divergence matrix is quite an easy task following \cite{tk81}.
Let's consider for example the element $x_{\mathsf{AC}}$; its derivative will be
\begin{equation}
\frac{dx_{\mathsf{AC}}}{dt}=\frac{d(q_{\mathsf{A}} q_{\mathsf{C}})}{dt}=q_{\mathsf{C}}\dot q_{\mathsf{A}} +q_{\mathsf{A}}\dot q_{\mathsf{C}}.\label{dt}
\end{equation}
It is worth giving a brief explication for this. 
We said that we are considering the two lineages at compositional equilibrium at the initial time,
so one would naturally say that $\dot q_{i} = 0$, and so the above equation.
Stating that we are at compositional equilibrium means that \textbf{sampling the whole considered sequence}
nucleotide frequencies $f_i$ don't change (apart from finite-size fluctuations). It does not mean that
there is no mutation at all on each site; had this been the case, there would be no evolution to study.
The probability for each nucleotide to mutate into another is given by the master equation, and this is why we
can write $x_{ij}$ as $q_i$ times $q_j$, take the derivative, and reexpress in terms of other $q_i q_j$ products,
I.E. other $\mathsf{X}$ entries.

An example of derivative would be, for example,
\begin{eqnarray}\label{dotadotc}
\dot q_{\mathsf{A}}=(dq_{\mathsf{C}}+bq_{\mathsf{G}}+aq_{\mathsf{T}})-(a+c+e)q_{\mathsf{A}}.
\end{eqnarray}
Substituting this and the analogue for $\dot q_{\mathsf{C}}$ in eq.(\ref{dt}) and doing the same for all $\mathsf{X}$ entries we obtain a
set of linear coupled first order differential equations which can be diagonalized and solved.

More detail on the derivation is reported in the appendix \ref{app1}.

\subsection{Reversibility}

Until now we have stated that it is possible to write the divergence matrix for this model, but it would be of no use because we could
never invert five expressions and obtain six independent rates as functions of the matrix entries. What can be done is to reduce 
the number of independent parameters by adding a relation between them. Many choices are possible. One could be, following \cite{tk81}, $a=f$.
Another possible choice is to make the model time reversible. We remember that time reversibility is satisfied when
\begin{equation}
p_{ij}f_{j}=p_{ji}f_{i} \qquad \forall i,j.
\end{equation}
where $p_{ij}$ are the entries of the evolutionary matrix and $f_{i}$ the equilibrium frequencies.
It is possible to demonstrate that this property is equivalent to the \textit{detailed balance} (see appendix \ref{app2}) which reads
\begin{equation}
r_{ij}f_{j}=r_{ji}f_{i} \qquad \forall i,j.
\end{equation}
In our model detailed balance holds if and only if
\begin{equation}
be=cd.
\end{equation}
This can be deduced by inspection of equilibrium frequencies expressions, or by a simpler rule \cite{luca},
reported here in appendix \ref{app3}.
A general version of reversible model has been studied by Yang \cite{ya94}, who pointed out its ability of fitting the data better than
other models. Gu and Li \cite{gu96} have shown its robustness against violation of time reversibility.

\section{\label{res}Results and discussion}

\subsection{Estimation of the substitution rates}
Due to the complexity of the expressions coming from this model, it is hard to think that one can
find an analytic way to invert them and express the rates as a function of the observables. Therefore we chose a statistic
way to perform this inversion, based on the $\chi^2$ test. We write the $\chi^2$ as
\begin{equation}
\chi^2 = \sum_{i,j} \frac{(\bar{x}_{i,j}-x_{i,j})^2}{\bar{x}_{i,j}} = \sum_{i,j} \frac{x_{i,j}^2}{\bar{x}_{i,j}} - 1.
\end{equation}

It is easily seen that this quantity is always non-negative, being zero when $\bar{x}_{i,j}=x_{i,j}$, I.E. when the model perfectly
fits the observations. Clearly, by performing a minimization on it we look at the same time for the best parameters.
In this contest trying to minimize the $\chi^2$ as a function of six parameters would outcome in a complete failure, the algorithm would
wander among the infinite number of equivalent solutions. Enforcing the reversibility makes the estimation possible, as it will be shown
below.

\subsection{A realistic example}

As an application, we started from the multiple alignment of rRNA sequences
used in \cite{Gouy89}. The observed divergence
matrix (unnormalized) between Xenopus and Homo is reported here below.
\mbox{}
\newline
\begin{center}
\begin{tabular}{c c c c c c}
\hline \hline
& & \multicolumn{4}{c}{ Xenopus }\\
\hline
& & A & T & G & C\\
& A & 647 & 1 & 17 & 2 \\
Homo & T & 3 & 523 & 11 & 18 \\
& G & 17 & 9 & 903 & 28 \\
& C & 8 & 21 & 25 & 691 \\
\hline \hline
\end{tabular}
\end{center}
\mbox{}
\newline

\begin{figure}[c]
\begin{center}
\includegraphics[width=7cm]{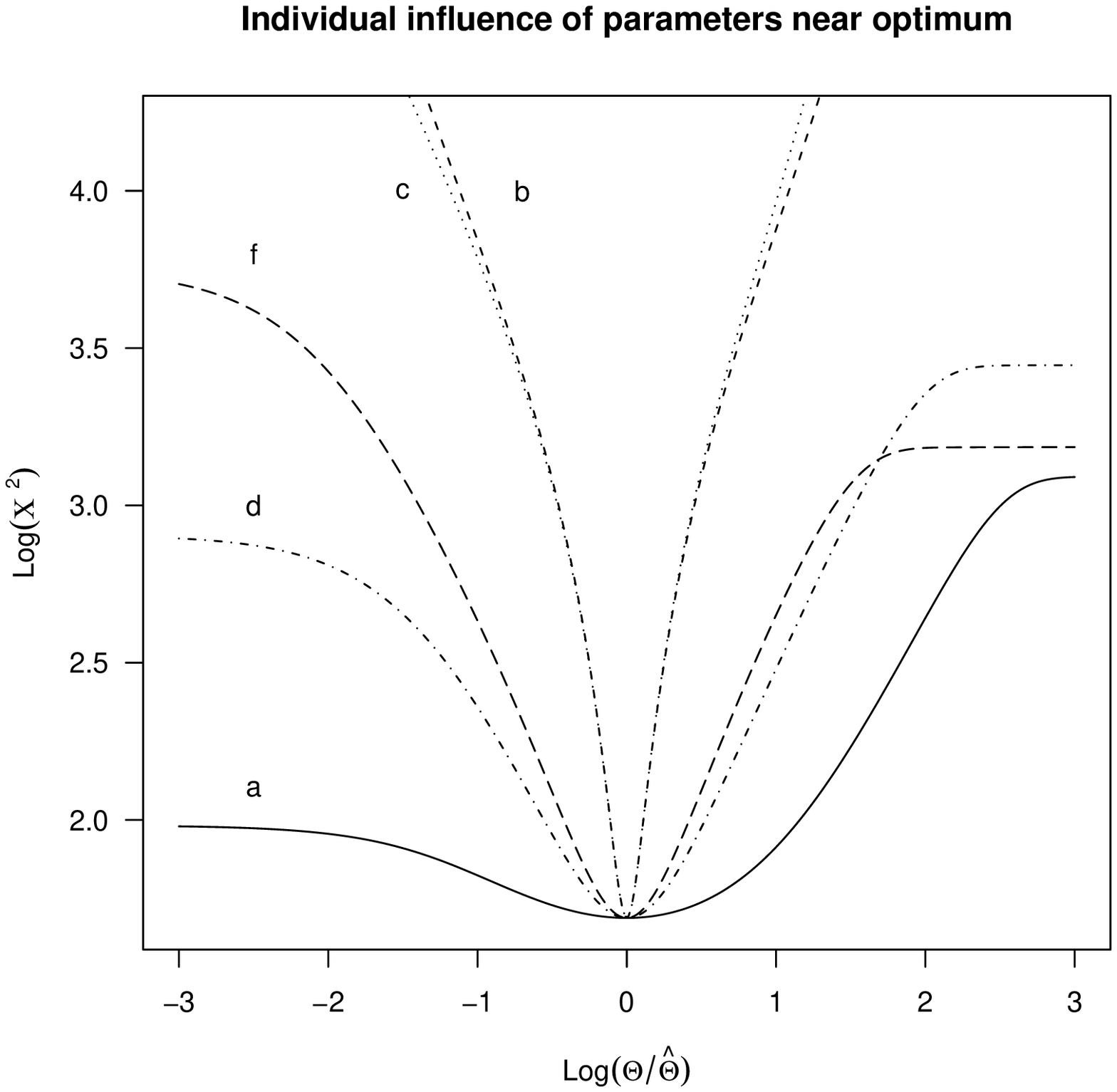}
\caption{\footnotesize{$\chi^2$ shaped as a minimum over 6 orders of magnitude.}}\label{paramfig}
\end{center}
\end{figure}

\begin{figure}[c]
\begin{center}
\includegraphics[width=7cm]{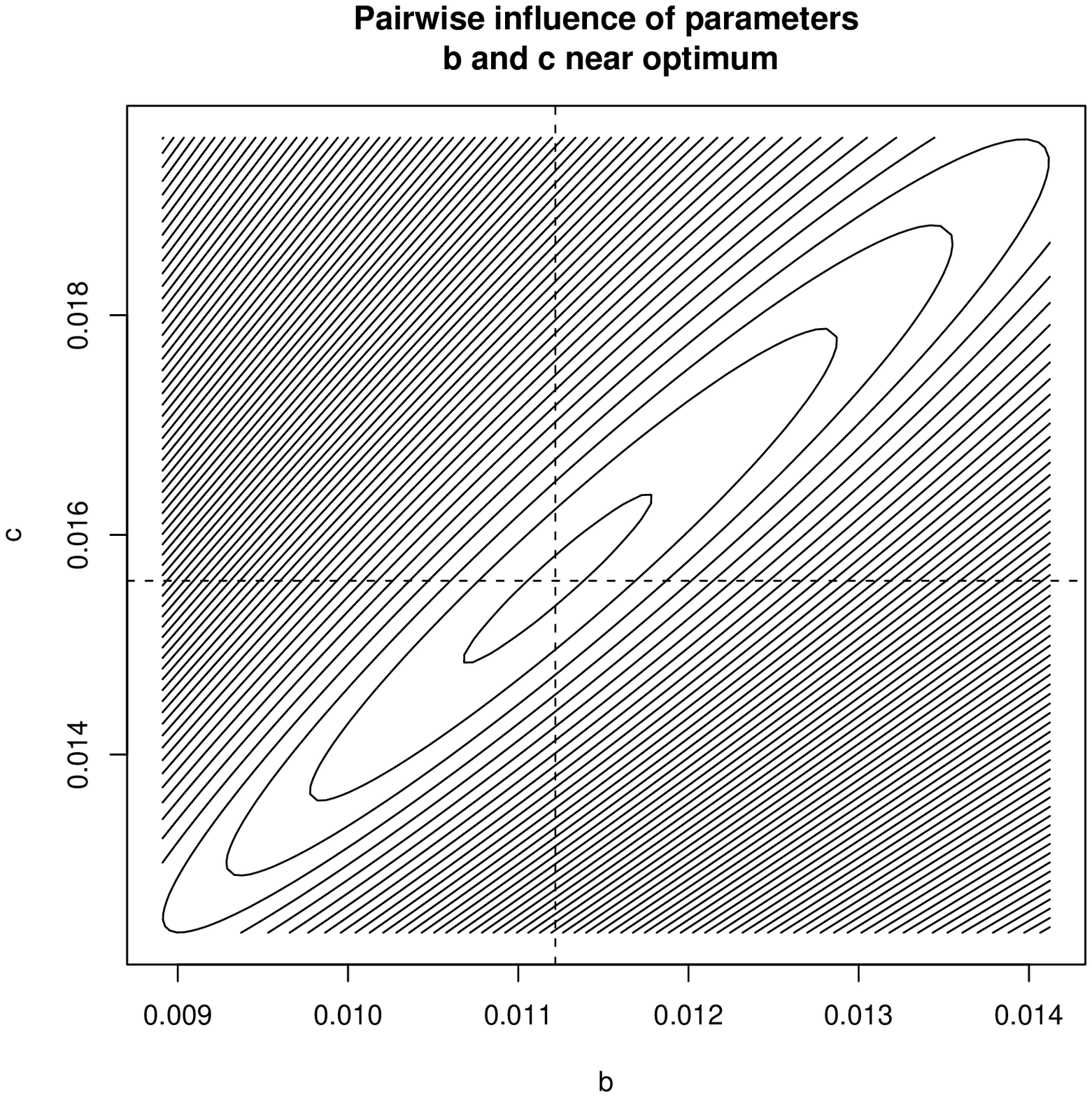}
\caption{\footnotesize{Near the optimal values for the parameters, only $b$ and $c$ show a structural correlation.}}\label{pairsfig}
\end{center}
\end{figure}

By changing parameter values over 6 magnitude orders we found that the $\chi^2$ criterion was well shaped with only one global minimum 
(fig. \ref{paramfig}).
A systematic exploration of all possible pairs of parameters showed that there were no strong structural correlations between parameters,
except between $b$ and $c$ (fig. \ref{pairsfig}). As a consequence, parameter values are easily estimated using standard non linear minimizing tools
(note that it is advisable to enforce parameter positivity during optimisation). This example showed that parameter can be estimated
in practice from a realistically sized dataset.

\subsection{Discussion}

The most general model of evolution at the DNA level has 12 parameters and this is
too much for practical purposes. If we try to simplify it by enforcing some parameter
to be equal, then the number of possible sub-models rapidly increases because many ways of doing it are possible.
At the opposite side we find the only model which requires all the parameters to be equal (JC).

It is clear that the number of published models in the literature doesn't cover all possible ones, and only those
coming from some biological or mathematical justifications have been explored.

Under {\it PR1 hypothesis}, we are dealing with {\it no strand-bias} models
whose most general form has 6 parameters.
We do not claim that models of this class are the best in any way, but that they are an interesting starting point.
An important property of these models is their convergence
towards {\it PR2 state} even if substitution rates are modified during
the course of evolution \cite{lolo99}. {\it PR2 state} is a strong assumption and strand asymmetry has been observed in
many cases. But, as {\it PR2} is usually
observed at a genome scale level \cite{lo95}, the hope is that, {\it on average},
with local deviations from {\it PR1 hypothesis} canceling out, this
class of model is not too bad an approximation.
The {\bf biological} motivation leading to the {\it no strand-bias}
models has an important {\bf mathematical} consequence,
so, if it is biologically reasonable to study these models, one must be aware of
the fact that the symmetry involved inexorably reduces the number of independent observations,
making the model mathematically intractable.

\subsection{Conclusion}
As we have shown in section \ref{counting} comparing the number of unknowns
to possible independent observations there is definitively no
hope to estimate the 6 parameters of the general form of the
{\it no strand-bias} model from pairwise DNA sequence comparisons.
There is no unique solution to a system of $M$ equations in $N > M$ unknowns,
in our case there is an infinite number of way to choose the six rates $a, b, c, d, e, f$ in order to satisfy the
five independent equations defining the matrix $\mathsf{X}$.
This result is extremely
unpleasant because it corresponds to the most common situation with
	experimental
data from present day DNA: fossil DNA data are scarce and from a relatively
recent past. We clearly need further simplifications.

We have exhibited here an example of a model, noted RNSB in figure 2, that
combines the properties of reversible models and {\it no strand-bias} models.
It is important to note that this model has still 5 parameters free because if the intersection
between the reversible model class and the {\it no strand-bias} class
were only --say--
3 parameter free models, there would not have been much flexibility left for further
research. We do not claim that this new RNSB model is the best intersection between
the two classes. We just claim that the RNSB model proves that it's possible to do so
with 5 free parameters, so that there is no bottleneck here for further theoretical
work on the parametric forms for this class of DNA substitution models.

\subsection*{Acknowledgements}
This contribution partly comes from the thesis OZ presented at Naples University in October 2002.
The authors thank warmly prof. Luca Peliti for connecting them during the \textsl{strapp 04} meeting (Dresden, Germany, July 5-10 2004).
OZ also because he was introduced by him to the beauties of biological systems.
They thank Manolo Gouy for kindly providing the multiple alignment of rRNA sequences and for many constructive suggestions.
The manuscript was also improved thanks to the comments from three anonymous reviewers.


\appendix
\section{Derivation of the divergence matrix}
\label{app1}

In order to obtain the expressions for the divergence matrix we define (following the notation introduced above)
\begin{eqnarray}\label{xyz}
X_{\pm}& \equiv 2S_{1} \pm 2Q_{1}\nonumber\\
Y_{\pm}& \equiv 2S_{2} \pm 2Q_{2}\\
Z_{\pm}& \equiv 4P     \pm 4R.\nonumber
\end{eqnarray}

These expressions reduce the problem to six first order ordinary coupled differential equations. This system is block-diagonal, 
can easily be inverted and its solution is:
\begin{eqnarray}\label{xyz+}
X_{+}&=&\omega[\omega+(1-\omega)e^{\lambda_{0}t}] \nonumber\\
Y_{+}&=&(1-\omega)(1-\omega+\omega e^{\lambda_{0}t})\\
Z_{+}&=&2\omega(1-\omega)(1-e^{\lambda_{0}t})\nonumber
\end{eqnarray}

and

\begin{eqnarray}\label{xyz-}
X_{-}&=&\frac{1}{g^{2}}\{2\beta[\alpha\omega-\beta(1-\omega)]e^{\lambda_{1}t}+\nonumber\\
   {}&{}& +[\zeta\omega+\beta^{2}(1-\omega)]e^{\lambda_{2}t}+\nonumber\\
   {}&{}& +[\eta\omega+\beta^{2}(1-\omega)]e^{\lambda_{3}t}\}\nonumber\\
Y_{-}&=&\frac{1}{g^{2}}\{-2\alpha[\alpha\omega-\beta(1-\omega)]e^{\lambda_{1}t}+\nonumber\\
   {}&{}&+[\alpha^{2}\omega+\eta(1-\omega)]e^{\lambda_{2}t}+\nonumber\\
   {}&{}& +[\alpha^{2}\omega+\zeta(1-\omega)]e^{\lambda_{3}t}\}\\
Z_{-}&=&\frac{1}{g^{2}}\{-2(\delta-\gamma)[\alpha\omega-\beta(1-\omega)]e^{\lambda_{1}t}+\nonumber\\
   {}&{}& +[\alpha(\delta-\gamma+g)\omega-\beta(\delta-\gamma-g)(1-\omega)]e^{\lambda_{2}t}+\nonumber\\
   {}&{}& +[\alpha(\delta-\gamma-g)\omega-\beta(\delta-\gamma+g)(1-\omega)]e^{\lambda_{3}t}\}\nonumber
\end{eqnarray}

where

\begin{eqnarray}\label{ab-}
          \alpha & \equiv & c-e\nonumber\\
          \beta  & \equiv & b-d\nonumber\\
          \gamma & \equiv & 2a+c+e\nonumber\\
          \delta & \equiv & b+d+2f\nonumber\\
     \omega & \equiv & 2f_{1}=2f_{A}=2f_{T}\nonumber\\
\lambda_{0} & \equiv & -2(b+c+d+e)\nonumber\\
\lambda_{1} & \equiv & -(2a+b+c+d+e+2f)\nonumber\\
\lambda_{2} & \equiv & \lambda_{1}+g\nonumber\\
\lambda_{3} & \equiv & \lambda_{1}-g\nonumber\\
          g & \equiv & \sqrt{(\delta-\gamma)^{2}+4\alpha\beta}\nonumber\\
      \zeta & \equiv & \frac{1}{2}(\delta-\gamma)(\delta-\gamma+g)+\alpha\beta\nonumber\\
       \eta & \equiv & \frac{1}{2}(\delta-\gamma)(\delta-\gamma-g)+\alpha\beta\nonumber
\end{eqnarray}

Combining all these, the entry for the divergence matrix are obtained.

\section{Reversibility and detailed balance}
\label{app2}
We will show here the equivalence between time reversibility and detailed balance.

\subsection{DETAILED BALANCE $\Rightarrow $ TIME REVERSIBILITY} 
Let's just remind that
$$
\mathsf{P}(t)=\exp\{\mathsf{R}t\},
$$
which can be developed as
\begin{equation}
\mathsf{P}(t)=\mathbb{I} +\mathsf{R}t + \frac{1}{2} \mathsf{R}^{2}t^{2} + \cdots,
\end{equation}
or
\begin{equation}
\label{ij}
p_{ij}=\delta_{ij} + r_{ij}t + \frac{1}{2} \sum_k r_{ik}r_{kj}t^{2} + \cdots
\end{equation}


Equation (\ref{ij}) can be also written as:
\begin{eqnarray}
p_{ij}&=&\delta_{ij}+\nonumber\\ 
{}&{}&+ \sum_{n=1}^{\infty}\frac{s_{ij}^{(n)}}{n!}t^n,
\end{eqnarray}
where
\begin{eqnarray}
s_{ij}^{(n)}&= &\sum_{k_{1}k_{2}\cdots k_{n-1}}r_{i,k_{1}}r_{k_{1},k_{2}}\cdots r_{k_{n-2},k_{n-1}} r_{k_{n-1},j}\nonumber\\ 
{}&{}&\quad \textrm{for}~ n\geq 2\nonumber\\
        {}&{}&{}\\
s_{ij}^{(n)}&= &r_{ij}, \qquad \qquad \qquad \qquad  \textrm{for}~ n=1. \nonumber
\end{eqnarray}
Now we will show that, if detailed balance is satisfied, then
\begin{equation}\label{s=}
s_{ij}^{(n)}f_{j}=s_{ji}^{(n)}f_{i}, \qquad \forall i,j,n.
\end{equation}
In fact, exploiting detailed balance,
\begin{eqnarray}
s_{ij}^{(n)}f_{j}=\sum_{k_{1}\cdots k_{n-1}}r_{i,k_{1}}\cdots r_{k_{n-1},j}f_{j}
\end{eqnarray}
becomes
\begin{eqnarray}
{}&\sum_{k_{1}\cdots k_{n-1}}r_{i,k_{1}}\cdots r_{j,k_{n-1}}f_{k_{n-1}}=\nonumber\\
=&\sum_{k_{1}\cdots k_{n-1}}r_{i,k_{1}}\cdots r_{k_{n-1},k_{n-2}}r_{j,k_{n-1}}f_{k_{n-2}}=\cdots\nonumber
\end{eqnarray}
and finally
\begin{equation}
\cdots=\sum_{k_{1}\cdots k_{n-1}}r_{k_{1},i}r_{k_2,k_1}\cdots r_{j,k_{n-1}}f_{i}.
\end{equation}
Reordering all the factors
\begin{eqnarray}\label{last}
\sum_{k_{1}\cdots k_{n-1}}r_{k_{1},i}r_{k_2,k_1}\cdots r_{j,k_{n-1}}f_{i}=\nonumber\\
\sum_{k_{1}\cdots k_{n-1}}r_{j,k_{n-1}}r_{k_{n-1},k_{n-2}}r_{k_{n-2},k_{n-3}}\cdots r_{k_1,i}f_{i} .
\end{eqnarray}
As the sum is performed on indices $k_{1} \cdots k_{n-1}$
the expression in (\ref{last}) is equal to $s_{ji}^{(n)}f_{i}$ for all $n \geq 2$.
So we have (\ref{s=}) for $n > 1$, and it is evident for $n=1$. Further, as
$\delta_{ij}f_{j}=\delta_{ji}f_{i}$, we obtain $p_{ij}f_{j}=p_{ji}f{i}$ \textit{Q. E. D.}

\subsection{DETAILED BALANCE $\Leftarrow $ TIME REVERSIBILITY}
Let's rewrite the formula
\begin{equation}\label{aga}
\frac{d\mathsf{P}(t)}{dt}=\mathsf{P}(t)\mathsf{R}; \quad \frac{dp_{ij}(t)}{dt}=\sum_{k}p_{ik}(t)r_{kj}.
\end{equation}
Let's compute the time derivative of $p_{ij}f_{j}$; if time reversibility holds it will be equal to the time derivative
of $p_{ji}f_{i}$.
From the formula (\ref{aga}), as equilibrium frequencies don't depend on time
\begin{equation}\label{dpdt2}
\frac{d}{dt}(p_{ij}(t)f_{j})=f_{j}\frac{dp_{ij}(t)}{dt}=\sum_{k}p_{ik}(t)r_{kj}f_{j}.
\end{equation}
But
$$
\frac{dp_{ij}(t)}{dt}=\sum_{k}r_{ik}p_{kj}(t),
$$
as $\mathsf{P}$ and $\mathsf{R}$ commute (evident from the solution).
The second expression in (\ref{dpdt2}) can be written as
\begin{equation}\label{dpdt3}
\sum_{k}p_{ik}(t)r_{kj}f_{j}=\sum_{k}r_{ik}p_{kj}(t)f_{j}.
\end{equation}
Because of the time reversibility the last expression in (\ref{dpdt3}) becomes
\begin{equation}\label{dpdt4}
\sum_{k}r_{ik}p_{kj}(t)f_{j}=\sum_{k}r_{ik}p_{jk}(t)f_{k}.
\end{equation}
Finally
\begin{equation}\label{dpdt5}
\frac{d}{dt}(p_{ji}(t)f_{i})=f_{i}\frac{dp_{ji}(t)}{dt}=\sum_{k}p_{jk}(t)r_{ki}f_{i}.
\end{equation}
Subtracting the (\ref{dpdt5}) from the (\ref{dpdt4}), which are equal, and keeping in evidence $p_{jk}(t)$ we finally obtain
\begin{equation}\label{dpdt6}
\sum_{k}p_{jk}(t)(r_{ik}f_{k}-r_{ki}f_{i})=0,
\end{equation}
and the detailed balance is satisfied \textit{Q. E. D.}

\section{Detailed balance: simple check}
\label{app3}
A nice property of detailed balance is that there exists a very easy way to state if it holds, 
even without calculating equilibrium frequencies.
Until now we have seen that the detailed balance is fulfilled when the equilibrium frequencies and the mutation rates (from which the 
former depend) cancel every term in the master equations.

Another way to check the detailed balance is to consider three states in the system and the rates connecting them. If the product of the 
three rates which takes from a state to itself ``clockwise'' is equal to that calculated ``counter-clockwise'', then the detailed balance 
holds. If we have three states $i, j, k$ then the above property will read
$$
r_{ik}r_{kj}r_{ji}=r_{ij}r_{jk}r_{ki}.
$$

\end{document}